\documentclass[twocolumn,aps,prd,unsortedaddress,%
superscriptaddress,a4paper,nofootinbib,balancelastpage,preprintnumbers,showpacs,floatfix]{revtex4}
\usepackage{latexsym,amsbsy,amsmath}
\renewcommand{\vec}[1]{\boldsymbol{#1}}
\renewcommand{\vec}[1]{\boldsymbol{#1}}
\def\ss#1#2{\boldsymbol{\sigma}_{#1}.\boldsymbol{\sigma}_{#2}}
\def\ll#1#2{\tilde{\lambda}_{#1}.\tilde{\lambda}_{#2}}
\def\llss#1#2{\tilde{\lambda}_{#1}.\tilde{\lambda}_{#2}\,\boldsymbol{\sigma}_{#1}.\boldsymbol{\sigma}_{#2}}
\usepackage{xcolor}
\begin{document}
\title{Search for doubly-heavy dibaryons in a quark model}
\author{J.~Vijande}
\email{javier.vijande@uv.es}
\affiliation{Departamento de F\'{\i}sica At\'{o}mica, Molecular y Nuclear, Universidad de Valencia (UV)
and IFIC (UV-CSIC), Valencia, Spain.}
\author{A. Valcarce}
\email{valcarce@usal.es}
\affiliation{Departamento de F{\'\i}sica Fundamental and IUFFyM,
Universidad de Salamanca, 37008 Salamanca, Spain}
\author{J.-M.~Richard}
\email{j-m.richard@ipnl.in2p3.fr}
\affiliation{Universit\'e de Lyon, Institut de Physique Nucl\'eaire de Lyon,
IN2P3-CNRS--UCBL,\\
4 rue Enrico Fermi, 69622  Villeurbanne, France}
\author{P.~Sorba}
\email{paul.sorba@lapth.cnrs.fr}
\affiliation{Laboratoire d'Annecy-le-Vieux de Physique Th\'eorique, CNRS-Universit\'e Grenoble-Alpes\\
 9 Chemin de Bellevue, 74940 Annecy-le-Vieux, France}
\date{
{\today}}
\begin{abstract}
We study the stability of hexaquark systems containing 
two heavy quarks and four light quarks within a simple quark model. 
No bound or metastable state is found. The reason stems on a 
delicate interplay between chromoelectric and chromomagnetic 
effects. Our calculation provides also information about 
anticharmed pentaquarks that are seemingly unbound in 
simple quark models.
\end{abstract}
\maketitle

\section{Introduction}
\label{se:intro}
Many interesting hadrons have been discovered recently with hidden heavy flavor, the $XYZ$ 
mesons and the LHCb pentaquarks. Some reviews are available, the latest ones including 
the hidden-charm pentaquarks and a few of them discussing also  
doubly-charm hadrons~\cite{Swanson:2006st,Chen:2016qju,Lebed:2016cna,Ali:2016gli,Hambrock:2013tpa,Nielsen:2009uh}.

The sector of doubly heavy-flavor will certainly call up a major experimental activity, in 
particular for a confirmation of the long-awaited doubly-charm baryons~\cite{Broadsky:2012rw} 
and for the search of doubly-charm mesons~\cite{Hyodo:2012pm,Che11} and other flavor-exotic 
states~\cite{Cho11}. 

In the case of $(\bar q \bar q QQ)$ with spin-parity $J^P=1^+$ and isospin $I=0$, where $Q$ stands 
for $c$ or $b$ and $q$ for a light quark, there is the fortunate cooperation of two effects. First, 
the chromoelectric interaction (CE), even if alone, gives stability below the $(\bar q Q)+(\bar q Q)$ 
threshold if the quark-to-antiquark mass ratio is large enough, as it takes advantage of the deeper 
binding of the $QQ$ pair~\cite{Ader:1981db,Zouzou:186qh,Heller:1986bt,Carlson:1988hh}. Second, the 
chromomagnetic interaction (CM) between the light quarks is also favorable~\cite{Semay:1994ht,Lee:2009rt}. 
Janc and Rosina predicted the stability of $(\bar u\bar d cc)$ using a quark model fitting 
ordinary hadrons~\cite{Jan04}. Their calculation was confirmed and improved by Vijande {\it et al.}~\cite{Vijande:2007rf}.
This configuration is also pointed out as a good candidate for a stable exotic in other approaches 
such as a $D D{}^*$ molecule~\cite{Manohar:1992nd,Ericson:1993wy,Tornqvist:1993ng,Ohkoda:2011vj,CarXX}, 
lattice QCD~\cite{Green:1998nt,Bali:2010xa,Bicudo:2015kna} or QCD sum rules~\cite{Dias:2011mi}. 

One hardly finds another configuration with both spin-independent and spin-dependent effects 
cooperating to privilege a collective multiquark state rather than a splitting into two hadrons. 
In this paper, we consider a multiquark system with two thresholds that might be nearly degenerate, 
and study to which extent the mixing of color and spin configurations can stabilize the multiquark. 
The system is $(qqqqQQ')$, where $Q$ or $Q'$ denotes a heavy quark and $q$ stands for a light quark. 
The dissociation thresholds are either of the $(qqq)+(qQQ')$ type, which benefits from 
the $QQ'$ CE interaction, or $(qqQ)+(qqQ')$ which can be shifted down by CM effects if 
$qq$ is in a spin singlet state. The question is whether $(qqqqQQ')$ can combine CE and 
CM dynamics coherently to build a bound state. 

In writing down the formalism and discussing the results, some other states will be evoked, 
such as the $H$ particle $(uuddss)$ in the limit of light $Q$ and $Q'$, or the pentaquark 
$(qqqq\bar Q)$ which is very similar in the limit where $Q$ and $Q'$ are clustered in a 
compact diquark. 

The paper is organized as follows. Section~\ref{se:model} contains the model and the method 
to calculate the relevant spin-color states and the matrix elements within this basis. 
In Sec.~\ref{se:var-cal} we present the variational calculation and its application in 
the case of states suspected to be either bound or weakly bound. The results are presented 
and discussed in Sec.~\ref{se:results}, while Sec.~\ref{se:outlook} is devoted to some 
conclusions and perspectives. 

\section{The model}
\label{se:model}
We consider $(qqqqQQ')$, where $Q$ and $Q'$ are heavy quarks which are different, hence
no Pauli constraints apply, but carry the same mass $M$ for simplicity. Giving $Q$ 
and $Q'$ different masses would not change our conclusions. We also take 
the SU(3)$_{\text F}$ limit in the light sector, with the same mass $m$ for $q=u,\,d$ or~$s$. 
For each baryon involved in the threshold and for the dibaryon, we search the ground state 
of the Hamiltonian $H=T+V$, where $T$ is the kinetic-energy operator and $V$ the interaction.
With $m= 0.4\,$GeV, $M=1.3\,$GeV, and a potential
\begin{multline}\label{eq:pot}
 V=-\frac{3}{16}\sum_{i<j} \ll{i}{j} \left( - \frac{a}{r_{ij}} + b\,r_{ij}\right.\\
 {}+ \left.\frac{c}{m_i\,m_j}\left(\frac{\mu}{\pi}\right)^{3/2}\exp(-\mu\,r_{ij}^2)\,\ss{i}{j}\right)\, ,
\end{multline}
where $a=0.4$, $b=0.2$, $c=2.0$, and $\mu=1.0$, in appropriate powers of GeV, one 
obtains a satisfactory account for the baryon masses entering 
the thresholds, both in the SU(3)$_\text{F}$ limit and with SU(3)$_\text{F}$  broken. A large negative constant is omitted in the above potential, 
but it can be disregarded, as it affects equally the thresholds and the multiquark energies. 
We do not elaborate here on the validity of a model with pairwise forces 
and color factors, which has been discussed already by several authors (see, e.g.,~\cite{Stanley:1980fe,Badalian:1987gg}). 
As it is, this is the simplest tool for such exploratory study. 

There is already an abundant literature on the wave functions of six-quark systems and 
the algebra of the spin, color, and spin-color operators entering the quark 
model~\cite{Hogaasen:1978xs,Mulders:1980vx,Park:2016cmg,Wang:1995kp,Leandri:1995zm,Leandri:1997ge,Pepin:1998ih}, with a careful account for the antisymmetrization constraints.
We thus restrict ourselves here to a brief summary of our notation. 
To construct the basis of color and spin states, we formally consider the system as a set of three 
two-quark subsystems, $(qq)(qq)(QQ')$, with color $\bar 3$ or $6$ and spin 0 or 1. We built the most general 
basis compatible with an overall color singlet and spin 0 state.
The requirements of antisymmetrization are strictly enforced 
for all states which are shown.

For an overall scalar, one can combine either three spins 0,
two spins 1 and one spin 0, or three spins 1, say
\begin{equation}\label{eq:spin-states}
\begin{gathered}
 S_1=(000)~,\quad
 S_2=(011)~,\quad
 S_3=(101)~, \\
 S_4=(110)~,\quad 
 S_5=(111)~.
\end{gathered}
\end{equation}

The simplest color singlet is $(\bar3\bar3\bar3)$ similar to any antibaryon made of 
three antiquarks. Another possibility consists of coupling two $\bar3$ diquarks 
into a color antisextet, and then to get an overall singlet with the third diquark 
being in a color sextet. The last possibility is to couple three $6$ diquarks into a 
singlet. In short,
\begin{equation}\label{eq:col-states}
\begin{gathered}
 C_1=(666)~,\quad
 C_2=(6\bar3\bar3)~,\quad  
 C_3=(\bar36\bar3)~, \\ 
 C_4=(\bar3\bar36)~,\quad
 C_5=(\bar3\bar3\bar3)~.
 \end{gathered}
\end{equation}
For the sake of cross-checking, the color states have been listed explicitly, 
and rearranged using the SU(3) Clebsch-Gordan coefficients given in~\cite{Alex:2010wi}.

The matrix elements of the spin, color and color-spin operators are obvious in this 
basis for the pairs $(1,2)$, $(3,4)$ or $(5,6)$. For the others, 
the crossing matrices corresponding to suited transpositions
have been used. For instance, the spin crossing matrix corresponding to 
\begin{equation} \label{eq:13-overlap}
 (12)(23)(34)\leftrightarrow (13)(24)(56)~,
\end{equation}
is  
\begin{equation}\label{eq:spin-X}
\frac{1}{2}\,
\begin{pmatrix}
  1 & 0 & 0 & \sqrt{3} & 0 \\
 0 & 1 & -1 & 0 & -\sqrt{2} \\
 0 & -1 & 1 & 0 & -\sqrt{2} \\
 \sqrt{3} & 0 & 0 & -1 & 0 \\
 0 & -\sqrt{2} & -\sqrt{2} & 0 & 0 \\
\end{pmatrix}
\end{equation} 
and gives access to $\ss{1}{3}$ and $\ss{2}{4}$. Its color analogue,
\begin{equation}\label{eq:colour-X}
\frac{1}{2}\,
\begin{pmatrix}
 -1 & 0 & 0 & \sqrt{3} & 0 \\
 0 & -1 & 1 & 0 & -\sqrt{2} \\
 0 & 1 & -1 & 0 & -\sqrt{2} \\
 \sqrt{3} & 0 & 0 & 1 & 0 \\
 0 & -\sqrt{2} & -\sqrt{2} & 0 & 0 \\
\end{pmatrix}
\end{equation}
allows for the calculation of $\ll{1}{3}$ and $\ll{2}{4}$.

Several checks can be made on the matrix elements $\ss{i}{j}$, $\ll{i}{j}$ and $\llss{i}{j}$. 
For instance, the Casimir operators such as $\sum \ss{i}{j}$ or  $\sum \ll{i}{j}$ depend only 
on the overall spin or color value. In the case of the $H$, the maximal value $\sum \llss{i}{j}=24$ 
is recovered, which exceeds the value $16$ corresponding to the $\Lambda\Lambda$ threshold, as 
first shown by Jaffe~\cite{Jaffe:1976yi}. Similarly, if the sum is restricted to the light sector, 
the maximal value $\sum \llss{i}{j}=16$ is obtained as in~\cite{Gignoux:1987cn,Lipkin:1987sk}, 
corresponding to more attraction than the value $8$ of the single baryon entering the lowest threshold. 

We note in Eq.~\eqref{eq:pot} the smearing of the spin-spin interaction, instead of a mere delta 
function when this term is treated at first order. Here the smearing parameter $\mu$ is the same for 
all pairs, unlike some more elaborate models, where it depends on the masses~\cite{Ono:1982ft}. 
For consistency, the stability is discussed with respect to the threshold computed within the same model. 

\section{Variational calculation}
\label{se:var-cal}
We solved the 6-body problem using a Gaussian expansion
\begin{equation}\label{eq:Gauss1}
 \phi(\tilde x) =\sum_i \gamma_i \exp[-(\tilde x^\dagger. A^{(i)}. \tilde x)/2]~,
\end{equation}
where $\tilde x=\{\vec x_1, \dots \vec x_5\}$ is a set of five Jacobi variables 
describing the relative motion, $\vec x_1=\vec r_2-\vec r_1$, etc., and $A^{(i)}$ a $5\times 5$ 
definite positive matrix. For a given choice of the matrices $A^{(i)}$, the weight 
factors $\gamma_i$ are given by a generalized eigenvalue problem. The overall variational 
energy is obtained by a numerical minimization over $A^{(i)}$.  

The calculation was started using a single spin-color channel, say
\begin{equation}\label{eq:Gauss2}
 \psi_{a,b} =\phi_{a,b} \,|C_a\rangle|S_b\rangle~,
\end{equation}
with $\phi_{a,b}$ given by Eq.~\eqref{eq:Gauss1}. It was later on extended to account 
for the coupling among the spin-color states, mainly due to the CM term, i.e., 
\begin{equation}\label{eq:Gauss3}
 \Psi =\sum_{a,b}\psi_{a,b}~,
\end{equation}
where the summation is extended to the states sharing the same symmetry properties. 
 
For estimating the energy of a deeply bound state, a straightforward strategy 
consists of choosing first a few diagonal matrices $A^{(i)}_{a,b}$ containing the 
range parameters, and then to add some non-diagonal terms and to increase the 
number of matrices. 

If binding does not show up, an alternative strategy consists of choosing 
Gaussians that describe two baryons times a relative function, for instance,
\begin{equation}\label{eq:Gauss4}
 \Psi=\phi_{123}\,\phi_{456} \sum_j \delta_j \,\exp(-\eta_j \,\vec r_{123-456}^2/2)~,
\end{equation}
where $\phi_{ijk}$ is a Gaussian approximation to the baryon containing the 
$\{i,j,k\}$ quarks and $\vec r_{123-456}$ links the centers of mass of the two 
baryons. Then the partitioning can be extended to other baryon-baryon configurations, 
say, in an obvious notation,
%
\begin{equation}\label{eq:Gauss5}
\Psi=\sum_{\substack{a=\{i,j,k\}\\ b=\{i',j',k'\}}}\hskip -9pt\phi_a\,\phi_b \sum_j \delta_j^{(a,b)} \,\exp(-\eta_j^{(a,b)} \,\vec r_{a-b}^2/2)~.
\end{equation}
This is similar to the method of Kamimura {\it et al.}~\cite{Hiyama:2003cu}, which 
has been applied successfully to a variety of few-body systems. 
\section{Results and discussion}
\label{se:results}
We first show the energies of the baryons constituting the thresholds in Table~\ref{tab1}.
\begin{table}[h!]
\caption{\label{tab1} Energy (in GeV) of the baryons involved in the thresholds within the model \eqref{eq:pot}. 
$\Sigma$ stands for a baryon where the first two quarks are in a spin 1 state, and 
$\Lambda$ in a spin 0 state.}
\begin{center}
\begin{ruledtabular}
\begin{tabular}{cccc}
             $qqQ (\Sigma)$ & $qqQ' (\Lambda)$ & $qqq (\Sigma)$ & $QQ'q(\Sigma)$ \\ \hline
             1.372          & 1.258           & 1.461          & 1.109          \\    
\end{tabular}
\end{ruledtabular}
\end{center}
\end{table}

We show in Table~\ref{tab2} the results
for the scalar state $J^P=0^+$ with isospin $I=1/2,3/2$,
that would be degenerate because the
potential in Eq.~(\ref{eq:pot}) does not depend on the 
total isospin.
This would stand, for example, for a flavor content 
$(uudscc)$\footnote{Other channels and flavor contents have been studied with similar results.}.
In this case thirteen different 
color-spin vectors are allowed by antisymmetry requirements, with
the notation of Eqs.~\eqref{eq:spin-states} and~\eqref{eq:col-states} they will be:
$C_1S_1$, $C_2S_1$, $C_3S_4$, $C_2S_2$, $C_3S_3$, $C_3S_5$, $C_1S_2$, $C_4S_3$,
$C_4S_4$, $C_4S_5$, $C_5S_3$, $C_5S_4$, and $C_5S_5$. The two thresholds allowed for the dissociation
of the $J^P=0^+$ six-quark state would have energies: $qqQ(\Sigma) + qqQ'(\Lambda)=$ 2.630 GeV and
$QQ'q(\Sigma) + qqq(\Sigma)=$ 2.570 GeV.
\begin{table}[ht!]
\caption{\label{tab2}Six-quark energies of the different color-spin vectors
contributing to the $J^P=0^+$ state, together with the coupled 
channel result and the energies of the allowed thresholds.}
\begin{center}
\begin{ruledtabular}
\begin{tabular}{cc}
Color-spin vector     & E (GeV) \\ \hline
$C_1S_1$  & 3.079 \\
$C_2S_1$  & 2.829 \\
$C_3S_4$  & 2.831 \\
$C_2S_2$  & 3.030 \\
$C_3S_3$  & 3.030 \\
$C_3S_5$  & 2.908 \\
$C_1S_2$  & 2.995 \\
$C_4S_3$  & 2.835 \\
$C_4S_4$  & 3.080 \\
$C_4S_5$  & 3.016 \\
$C_5S_3$  & 2.891 \\
$C_5S_4$  & 2.997 \\
$C_5S_5$  & 3.034 \\
Coupled   & 2.767 \\ \hline
Thresholds & 2.570, 2.630 \\
\end{tabular}
\end{ruledtabular}
\end{center}
\end{table}
We also give in Table~\ref{tab3} the probabilities of the different channels contributing
to the coupled channel calculation (those that are not listed have probabilities smaller
than $10^{-6}$).
\begin{table}[ht!]
\caption{\label{tab3}Probabilities of the different six-body channels contributing to the 
$J^P=0^+$ six-quark state.}
\begin{center}
\begin{ruledtabular}
\begin{tabular}{ccccc}
 Channel       & $C_1S_2$ & $C_2S_1$ & $C_3S_4$ & $C_4S_3$ \\ \hline
 Probability   & 0.004    & 0.539    & 0.456    & 0.001     \\    
\end{tabular}
\end{ruledtabular}
\end{center}
\end{table}

The calculations using the variational wave function~\eqref{eq:Gauss2} for a single channel, 
and~\eqref{eq:Gauss3} for the case of coupled channels always give values above the threshold, which 
go down very slowly when the Gaussian basis is augmented.
As already said, this is the sign of either 
the absence of a bound state, or, at most, of a very tiny binding. This is confirmed by the use of the 
alternative bases, Eq.~\eqref{eq:Gauss4} or Eq.~\eqref{eq:Gauss5}, where one always finds a 6-body energy 
equal to the sum of the two baryon energies, obtained in the approximation of the Gaussian expansion 
$\phi_{ijk}$ and $\phi_{i'j'k'}$. This means that neither the residual color-singlet exchange between  
the two clusters, nor the coupling of the different baryon-baryon thresholds is sufficient to bind the system. 

We have checked that in the infinite mass limit for the mass of the heavy quarks the system gets binding
with respect to the upper threshold, $(qqQ)+(qqQ')$, but it is always above the lowest one $(QQ'q)+(qqq)$.
For example, for $M=10$ GeV and $m=0.4$ GeV we get 2.326 GeV for the energy of the six-quark state
in the coupled channel calculation,
while the thresholds come given by $E(QQ'q)+E(qqq)=$ 2.162 GeV and $E(qqQ)+E(qqQ')=$ 2.477 GeV. 
The six-quark state, that it is now in between the two thresholds, is described by the same color-spin vectors
shown in Table~\ref{tab3}, $C_2S_1$ and $C_3S_4$, where the two-heavy quarks are in a
$\bar 3$ color state, see Eq.~\eqref{eq:col-states}, that would split into the lowest threshold. 
In other words, the two-heavy quarks
control the mass of the six-body state in the infinite mass limit. As mentioned 
above, by making use of the variational wave function of Eq.~\eqref{eq:Gauss4} or Eq.~\eqref{eq:Gauss5}
one obtains exactly the two-baryon mass in the six-body calculation.

We now try to explain why these results are plausible. 
For the CM part, the subject is already well documented, with the discussions around the $H$ dibaryon or 
the 1987-vintage pentaquark. See, for instance,~\cite{Oka:1983ku,Rosner:1985yh,Karl:1987cg}. The effects 
of SU(3)$_{\rm F}$ breaking, a different mass for the strange quark, tends to spoil the promises of binding based on 
the sole spin-color algebra, and, more important, the short-range correlation factors (the expectation 
values of $\exp(-\mu\,r_{ij}^2)$ in our model) are significantly smaller in a multiquark than in baryons. 

As for the CE part, a superficial analysis would argue that, as soon as $-\sum \ll{i}{j}$ is locked to 
$16$ in any spin-color channel $|C_a\rangle|S_b\rangle$, the CE part of the binding will remain basically 
untouched, independent of the combination of the $|C_a\rangle|S_b\rangle$ dictated by the CM part. However, this 
is not the case. For equal masses, the deepest CE binding is obtained when the distribution of CE strength 
factors $\{-\ll{i}{j}\}$ is the most asymmetric~\cite{Ric11}, which favors the threshold against a compact 
multiquark. For a mass distribution such as $(qqqqQQ')$, CE dynamics favors the $QQ'$ two-quark state being in a 
color $\bar 3$ state. Once this is enforced, the best CE energy is obtained when the $Qq$ and $Q'q$ pairs 
receive the largest strength, and they come with a larger reduced mass than $qq$. This can be checked 
explicitly in a simple solvable model with an interaction proportional to $ -\ll{i}{j}\,r_{ij}^2$. However, CM effects 
are optimized when the light sector receives the largest color strengths. Hence, there is somewhat a 
conflict between CE and CM effects, and this explains the lack of bound states in our model.

\section{Summary and outlook}\label{se:outlook}
In this paper, we have used a simple quark model with CE and CM components to search for possible bound states 
of $(qqqqQQ')$ configurations below their lowest threshold. The answer is negative: no bound state is found, 
nor any metastable state in the continuum with a mass below the highest threshold and a suppressed decay to the 
lowest threshold. 

Our calculation provides also some information about the anticharmed pentaquark, or beauty analog $P=(qqqq \bar Q)$. 
In the limit of large $Q$ and $Q'$ masses, the $QQ'$ pair in $(qqqqQQ')$ behaves as a single antiquark. 
>From our results, $(qqqq \bar Q)$ is seemingly unbound in simple quark models, while a simple CM counting 
suggests that this configuration is bound~\cite{Gignoux:1987cn,Lipkin:1987sk}. While the $H$ has been much 
studied, in particular within lattice QCD~\cite{Bea10,Ino10,Bea11,Ino12,Yamazaki:2015nka}, 
the $P$ has received less attention.

It is worth to emphasize how our study illustrates the difficulty to get multiquark 
bound states in constituent quark models. Other 
approaches have suggested some ways out, such as:
\begin{itemize}

\item Another spin-dependent part for our potential~\eqref{eq:pot}. For instance, 
a chiral quark model was considered in~\cite{Pepin:1998ih} for the $H$ and for $(uuddsQ)$, 
but no bound state was found.

\item Multibody potentials, generalizing the $Y$-shape potential for baryons, provide more attraction 
than the color-additive model~\eqref{eq:pot}~\cite{Vijande:2011im}, but in the minimization of the flux 
tube configurations, several color states are mixed, and this is delicate in the case of identical quarks, 
where color is constrained by the requirement of antisymmetry~\cite{Vijande:2013qr}.
 
\item Diquarks, whose clustering is motivated by CM effects but not really demonstrated, might lead to 
dibaryon states~\cite{Fredriksson:1981mh,Maiani:2015iaa}.
 
\item String dynamics {\`a} la Rossi-Veneziano suggests the existence of states containing more junctions 
that the lowest thresholds, and thus metastable, as the internal annihilation of junctions is suppressed 
by an extension of the Zweig rule~\cite{Rossi:2016szw}. 
 
\item Molecular dynamics, doubly-heavy dibaryons have been also recently approached in a molecular 
picture~\cite{Lee11,Meg11,Liz12,Oka13,Hua14,Car15}. The main
motivation of these studies originates from the reduction of the kinetic energy due
to large reduced mass as compared to systems made of light baryons.
However, such molecular states that have been intriguing objects of 
investigations and speculations for many years, are usually the concatenation of
several effects and not just a fairly attractive interaction. The coupling 
between close channels or the contribution of non-central forces used to 
play a key role for their existence. 
When comparing to similar problems in the strange sector the mass difference 
between the two competing channels $(qqQ)+(qqQ')$ and $(qqq)+(qQQ')$ increases,
making the coupled channel effect less important. Thus, without the strong 
transition potentials reported in the QDCSM model of Ref.~\cite{Hua14} or 
the strong tensor couplings occurring in the hadronic one-pion exchange 
models of Refs.~\cite{Meg11,Liz12}, it seems difficult to get a molecular 
bound state of two heavy baryons, as has been recently reported in Ref.~\cite{Car15}.
\end{itemize}

On the experimental side, the search for doubly-charm dibaryons can be made together
with the search for doubly-charm baryons~\cite{Chn14,Che14,Zhe16} and 
doubly-charm exotic mesons~\cite{Hyodo:2012pm,Cho11,Che11} as they share some triggers.

\acknowledgments
We benefited from fruitful discussions 
with Emiko Hiyama and Makoto Oka.
This work has been partially funded 
by Ministerio de Educaci\'on y Ciencia and EU FEDER under 
Contracts No. FPA2013-47443 and FPA2015-69714-REDT,
by Junta de Castilla y Le\'on under Contract No. SA041U16,
and by Generalitat Valenciana PrometeoII/2014/066.
A.V. is thankful for financial support from the 
Programa Propio XIII of the University of Salamanca.

\end{document}